\documentclass[12pt,twoside]{article}
\usepackage{xrb2007}
\usepackage{graphicx}
\usepackage{graphics}
\usepackage{psfig}
\usepackage{epsf}
\usepackage{lscape}
\def \inte {{$INTEGRAL$}}
\def \xmm {{\em XMM--Newton}}

\def \src {\mbox{IGR~J11215--5952}}
\def \degmark{^\circ}

\def \hcm {\hbox {\ifmmode $ atom cm$^{-2}\else atom cm$^{-2}$\fi}}

\def \mdot {\dot{M}_{w}}

\begin{document}

\session{Obscured XRBs and INTEGRAL Sources}

\shortauthor{Sidoli}
\shorttitle{A new explanation for the SFXTs outbursts}

\title{A new explanation for the Supergiant Fast X--ray Transients outbursts}
\author{Lara Sidoli}\affil{INAF, Istituto di Astrofisica Spaziale e Fisica Cosmica, Via E.\ Bassini 15,   I-20133 Milano, Italy}\author{Patrizia Romano}\affil{INAF, Osservatorio Astronomico di Brera, Via E.\ Bianchi 46, I-23807 Merate (LC), Italy}\author{Sandro Mereghetti, Ada Paizis, Stefano Vercellone}\affil{INAF, Istituto di Astrofisica Spaziale e Fisica Cosmica, Via E.\ Bassini 15,   I-20133 Milano,  Italy} \author{Vanessa Mangano}\affil{INAF, Istituto di Astrofisica Spaziale e Fisica Cosmica, Via U.\ La Malfa 153, I-90146 Palermo, Italy } \author{Diego G\"{o}tz}\affil{CEA-Saclay, DAPNIA/Service d'Astrophysique, F-91191 Gif-sur-Yvette Cedex, France}

\begin{abstract}
The physical mechanism responsible for the short outbursts in a recently
recognized class of High Mass X-ray Binaries, the Supergiant Fast X-ray
Transients (SFXTs), is still unknown.
Recent observations performed with $Swift/XRT$, \xmm\ and \inte\ of
the 2007 outburst from IGRJ11215$-$5952, the only SFXT known to exhibit
periodic outbursts, suggest a new explanation for the outburst mechanism
in this class of transients; the outbursts could be linked to the possible presence of a second
wind component in the supergiant companion, in the form of an equatorial wind.
The applicability of the model to the short outburst durations of all
other Supergiant Fast X-ray Transients, where a clear periodicity in the
outbursts has not been found yet, is discussed. The scenario we are proposing
also includes the persistently accreting supergiant High Mass X--ray Binaries.
\end{abstract}

\section{Introduction: the SFXTs properties}

Supergiant Fast X--ray Transients (hereafter SFXTs;  Smith et al. 2006a) 
are a new class of
hard X--ray sources mostly discovered by the $INTEGRAL$ satellite 
(Negueruela et al. 2005a, Sguera et al. 2005).
They are transient sources which seem to
emit X--rays only during ``short'' outbursts (few hours, as observed with $INTEGRAL$ or $RXTE$) and 
their optical counterparts are blue supergiant stars.
Their X--ray spectra are similar to those of accreting pulsars, thus it is likely that
they are High Mass X--ray Binaries (HMXBs) hosting neutron stars. 
In two SFXTs (among about twenty sources,
comprising  candidate SFXTs) X--ray pulsations 
have been indeed observed (IGR J18410-0535/AX J1841.0-0536, 
P$_{\rm spin}$=4.74~s, Bamba et al. 2001; 
IGR~J11215--5952, P$_{\rm spin}=186.78\pm0.3$\,s, Smith et al. 2006b, Swank et al. 2007).

SFXTs reach X--ray 
luminosities up to a few 10$^{36}$~erg~s$^{-1}$, 
while the quiescent level 
($\sim$10$^{32}$~erg~s$^{-1}$) has been
observed only in  IGR~J17391$-$3021/XTE~J1739$-$302 (Smith et al. 2006a; Sakano et al. 2002), 
IGR~J17544$-$2619 (in't Zand 2005) and probably 
IGR~J18410$-$0535/AX~J1841.0$-$0536 (Halpern et al. 2004; Bamba et al. 2001).

It is important to note that 
none of the  SFXT sources has ever been caught
to undergo a transition from
quiescence to outburst 
and then back to quiescence in a few hours. 
The quiescent emission had always been observed
well far away from the outbursts, except in one case:
only in't Zand (2005) did observe the transition from quiescence to outburst with $Chandra$ 
(in IGR~J17544$-$2619),
but the observation finished before the start of the declining phase to quiescence.
Thus the real duration of this outburst could not be measured.
The so-called ``short'' duration (a few hours) of the outbursts from SFXTs is indeed based 
on  observations with instruments not sensitive enough
to detect the quiescence level. The instruments onboard $RXTE$ and $INTEGRAL$ 
could  only observe the brightest fast  
flaring activity (lasting a few hours, less than one day) reaching a few 10$^{36}$~erg~s$^{-1}$. 

Hence, the definition of SFXTs as transient sources displaying ``short'' X--ray 
outbursts lasting only a few hours is strongly biased.
This has been observationally demonstrated by our recent 
deep campaign with $Swift/XRT$  (Romano et al. 2007, hereafter Paper~II)
of the outburst
from the unique SFXT displaying ``periodic outbursts'', 
IGR~J11215--5952 (Sidoli et al. 2006, hereafter Paper~I). 
These very sensitive observations 
showed that the
accretion phase in SFXTs lasts longer than what previously thought: a few days instead of only hours.

With these new observations at hand, we report on an alternative model to explain the outbursts from this new class of sources,
based on $Swift/XRT$ monitoring observations of IGR~J11215--5952 during the last two outbursts 
(starting on February 9 and July 24, 2007; Sidoli et al. 2007, hereafter Paper~III).

\section{Swift/XRT observations of IGR~J11215--5952}

\src\ is an X--ray transient  discovered by 
\inte\  during a fast outburst in  April 2005 (Lubinski et al. 2005).
The optical counterpart is  a B1 supergiant,
HD~306414, located at a distance of 6--8~kpc 
(Negueruela et al. 2005b, Masetti et al. 2006, Steeghs et al. 2006).
From the analysis of \inte\ observations of the source field, we discovered (Paper~I)
that the outbursts are equally spaced by $\sim$330~days (although
a half of this period could not be excluded, due to a lack of observations).
This periodicity
was later confirmed in March 2006 (outburst after 329 days; Smith et al. 2006c) 
during a monitoring with $RXTE/PCA$,
and was related in a natural way to the orbital period of the system, with
the outbursts triggered at (or near to) the periastron passage (Paper~I).

Based on this known periodicity, a new outburst was expected for 2007 February 9 and 
we planned a  monitoring campaign with $Swift/XRT$, starting on  2007 February 4 
(Romano et al. 2007b).
A second monitoring campaign was performed with $Swift/XRT$ in July 2007, in order to
monitor the quiescent level and the epoch of the supposed apastron passage (based
on the 329 days period; Romano et al. 2007c). These observations led to the
detection of a new unexpected outburst starting on 2007, July 24, which reached roughly the
same flux as during the February 2007 outburst (Paper~III).
Details of the $Swift/XRT$ data analysis and spectral/timing results 
are reported in Paper~II and Paper~III. 
Here we concentrate on the shape of the X--ray lightcurve in order to understand
the physical mechanism which produces the outbursts.


\section{A new model for the  outburst mechanism in SFXTs}

\begin{figure*}[th!]
\includegraphics[angle=-90,width=10.5cm]{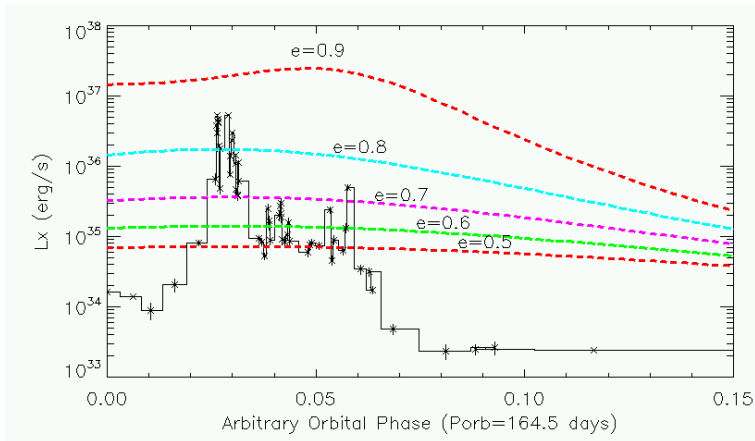}
 		\caption{Comparison of the \src\ lightcurve observed with $Swift/XRT$ during
the February 2007 outburst, with the 
          X--ray luminosity calculated from Bondi-Hoyle accretion from the spherically symmetric
wind of the B1-type supergiant 
companion, for different binary system eccentricities. An orbital
period of 164.5~days is assumed. The orbital phase is arbitrary. 
We assumed a beta-law for the supergiant wind, with an exponent $\beta$=1,
a stellar mass of 39~M$_\odot$, a radius of 42~R$_\odot$, 
a wind terminal velocity of 1200~km~s$^{-1}$ and a wind mass loss of 
3.7$\times$10$^{-6}$~M$_\odot$~yrs$^{-1}$.
}
                \label{fig:ecc}
\end{figure*}

The IGR~J11215--5952 lightcurve observed during the February 2007 outburst 
represents
the most complete set of observations of a SFXT outburst 
(Fig.~\ref{fig:ecc}, black curve).
The first important result of these observations is that the 
whole outburst phase lasts longer than what previously thought, based on less sensitive
instruments: a few days, instead of a few hours.
Only the brightest part of the outburst is short (lasts less than 1 day) and
would have been seen by the INTEGRAL instruments. 
Intense flaring activity is also present, both during the bright peak and  the declining
phase of the outburst, with each single flare lasting minutes or a few hours.

\vspace{-1.5cm}

It is natural to associate the clock responsible for  the  outbursts with 
the orbital periodicity of the binary system. 
Since \src\ displayed a new outburst after about a half of the 329~days period 
(Paper~III; Romano et al. 2007c), it is possible
that 164.5 days is indeed the real orbital period which escaped detection up to now.
In both cases (P$_{\rm orb}$=329 days or 164.5 days), the system is a wide binary where
the blue supergiant does not fill its Roche lobe, and the system is very likely wind-fed.
Applying the Bondi-Hoyle wind
accretion scenario, where the neutron star accretes from the wind of the supergiant 
at different rates depending on the wind density and relative velocity along the orbit,
and
assuming reasonable parameters for the B-supergiant, we obtain that the observed X--ray lightcurve 
is always too narrow and steep to be explained with accretion from a spherically symmetric wind,
even adopting extreme eccentricities for the binary system (see Fig.~\ref{fig:ecc}).

\begin{figure*}[th!]
\vbox{
\includegraphics[angle=0,height=8.4cm]{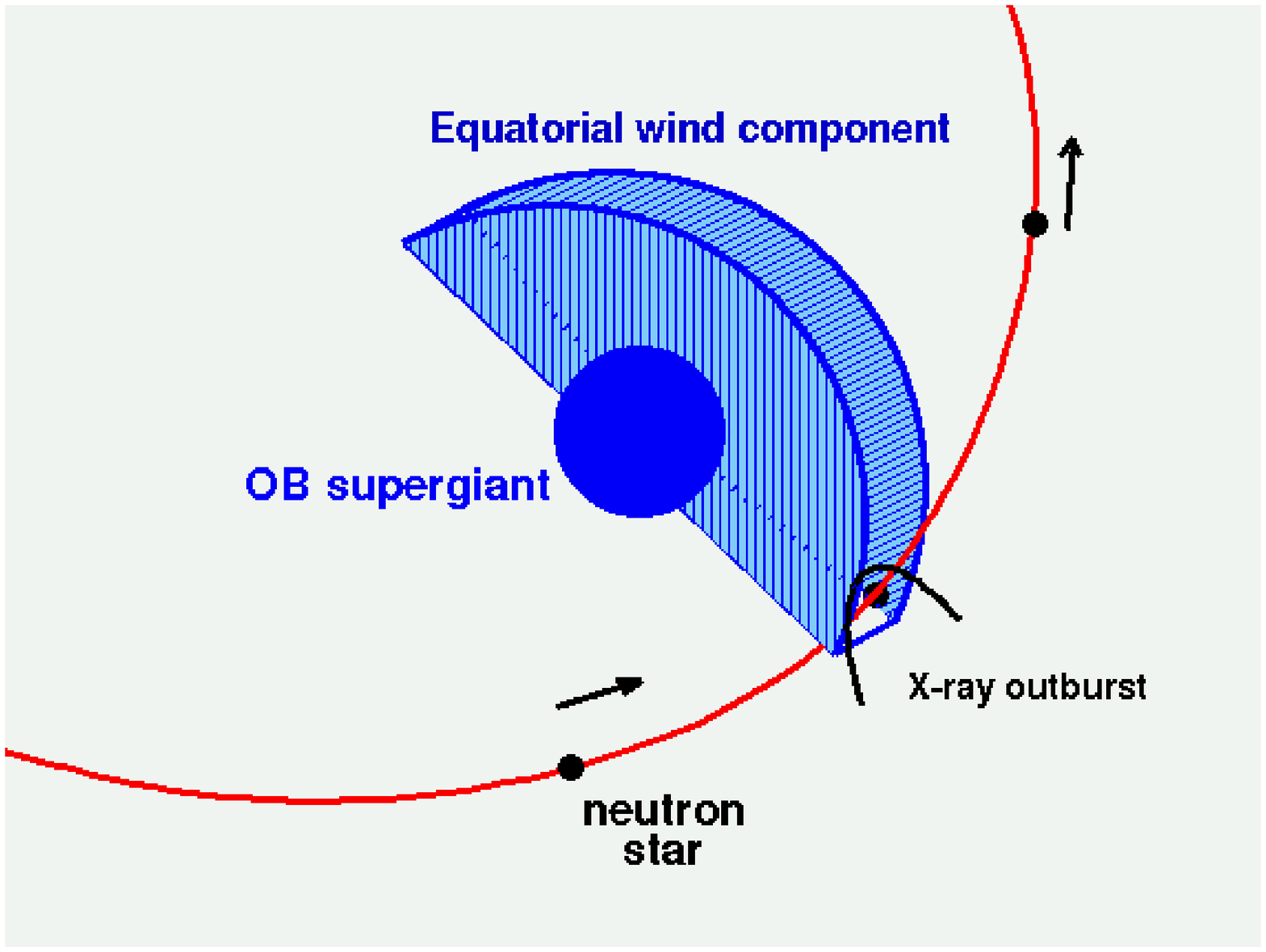}
\hspace{0.1cm}
\includegraphics[angle=0,height=8.4cm]{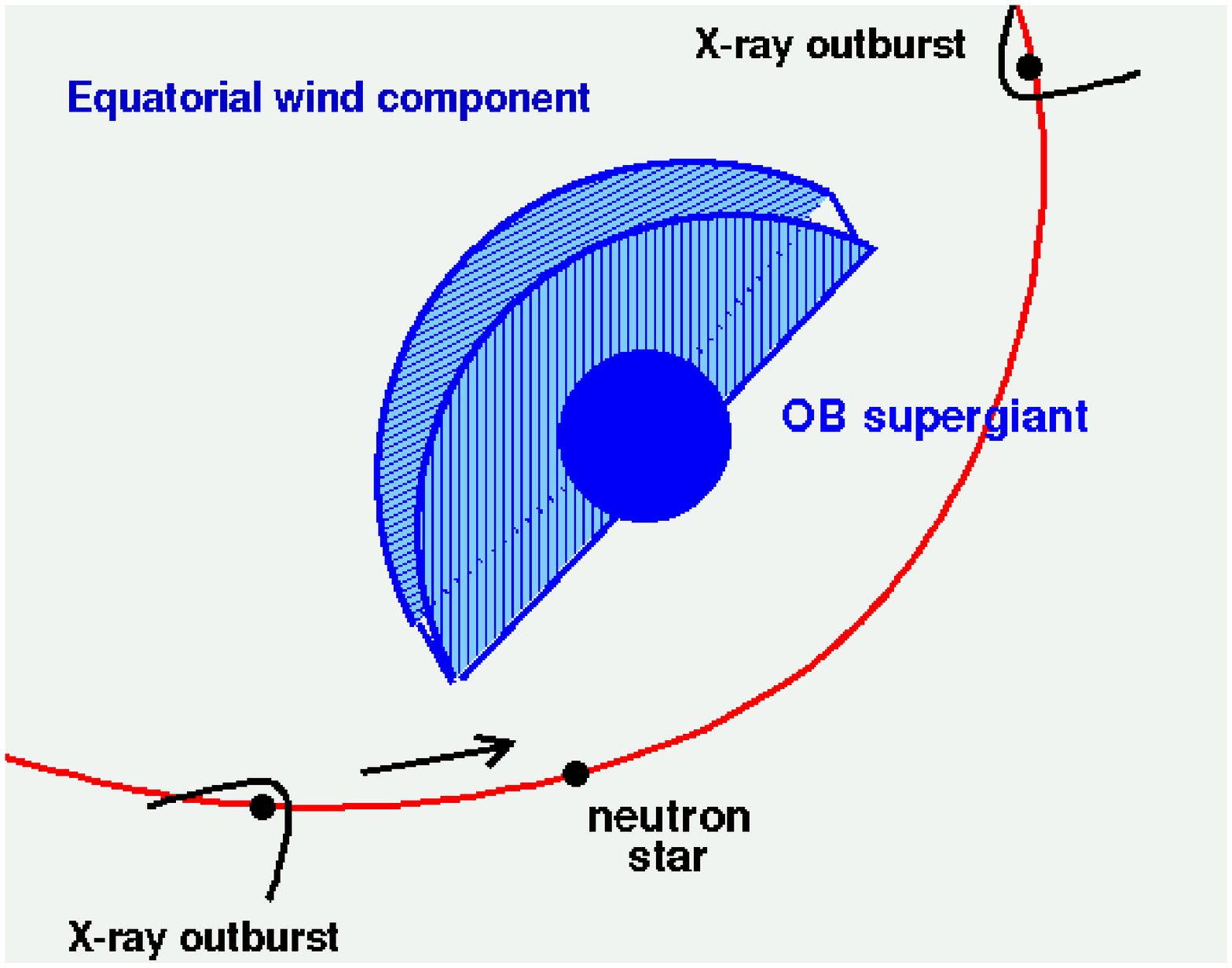}}
\vspace{-2.cm}
\caption[]{{\em Left panel:} sketch of the proposed geometry for the equatorial wind component
and the outburst mechanism in \src. {\em Right panel:} sketch of the proposed geometry
for a SFXT in general (see text).
}
\label{fig:geom}
\end{figure*}

This result led us to suggest that the wind from the B supergiant is not spherically symmetric.
The alternative viable explanation we propose for the sharpness of the observed X--ray lightcurve
is  that in \src\ 
the supergiant wind has a second  component (besides the polar
spherically symmetric one), in the form of an ``equatorial disk'', inclined with respect to
the orbital plane (see Fig.~\ref{fig:geom} for an artistic view of the geometry of the system).
The short outburst  is then produced when the neutron star crosses this equatorial wind component, 
denser and slower than the polar one. 
Deviations from spherical symmetry in hot massive star winds
are also suggested by optical observations (e.g. Prinja 1990,  Prinja et al. 2002) 
and the presence of 
equatorial disk components, denser and slower with respect to the polar wind,
also results from simulations  (ud-Doula et al. 2006).

The thickness {\em h} of the  densest part of this 
supergiant equatorial wind
can be calculated from the duration of the brightest part of the outburst (which
lasts less than 1~day, time needed for the neutron star to cross it) and the
neutron star velocity, 100--200~km~s$^{-1}$: 
{\em h}$\sim$(0.8--1.7)$\times$sin$(\theta)$$\times$10$^{12}$~cm,
where $\theta$ is the inclination angle of the equatorial wind 
with respect to the
orbital plane (with $\theta$=90$\degmark$ if the disk 
is perpendicular to the orbital plane).

The model we are proposing can explain also the short flares
from all the other SFXTs
where a clear periodicity in the outbursts recurrence has not been found yet, 
if a different geometry of the equatorial wind component with respect to
the orbital plane is assumed: in \src, where the outbursts are equally spaced and 
occur with a fixed periodicity, 
the
inclined equatorial disk wind component should intersect 
the neutron star at the periastron
(or very close to it, see the left panel in Fig.~\ref{fig:geom}) and can intersect the 
neutron star orbit once or twice depending
both on the extension of the wind disk and on the orbital eccentricity.
Instead, it is possible that 
in the other SFXTs the inclined disk wind intersects twice 
a wide and highly eccentric orbit, not at the periastron 
(see the right panel in Fig.~\ref{fig:geom}), 
leading to a double
periodicity (one shorter than the other) which has not been found yet {\em only} because of 
a  lack of a continuous monitoring.
This model can also explain  the X--ray emission from 
the persistently accreting HMXBs, 
if we admit that in this case the neutron star is always moving 
inside the equatorial wind component
which lies on the orbital plane.

In this framework, the  sharp X--ray lightcurve observed from \src\ can be modelled with different
wind parameters (for both polar and equatorial components) 
depending on the orbital period (164.5 days or 329 days)
and the eccentricity of the binary.
We assume a blue supergiant with a mass of  
39~M$_\odot$ and radius of 42~R$_\odot$, and a  polar wind component  with
a terminal velocity of 1800~km~s$^{-1}$.
The X--ray lightcurve observed with $Swift/XRT$ is better reproduced assuming
a ``polar wind'' mass loss rate of 
5$\times$10$^{-6}$~M$_\odot$~yrs$^{-1}$ (for a P$_{\rm orb}$ of 164.5 days and 
an eccentricity of 0.4) and
9$\times$10$^{-7}$ ~M$_\odot$~yrs$^{-1}$ (for a P$_{\rm orb}$ of 329 days and 
a circular orbit, which is required
by the fact that the two consecutive outbursts from \src\ reached roughly 
the same peak flux). 
The equatorial wind component should have  a 
variable velocity ranging from 750~km~s$^{-1}$ to 1400~km~s$^{-1}$ (for P$_{\rm orb}$=164.5 days),
and from 850~km~s$^{-1}$ to 1600~km~s$^{-1}$  (for  P$_{\rm orb}$=329~days), and a density about 100 times
higher than the polar wind component.

Note however that since the X--ray luminosity expected for the wind accretion 
is proportional to $\mdot$$v_{rel}$$^{-4}$ (where $\mdot$ is the wind mass loss rate,
and $v_{rel}$ is the relative velocity of the wind with respect to the neutron
star), different combinations of wind density and velocity
in the equatorial component can reproduce the X--ray lightcurve as well.

In conclusion, in our model we explain the short recurrent flares 
if the neutron star intersects an inclined 
equatorial 
wind component (once or twice) during its orbit.
A different particular geometry and inclination of this equatorial wind 
with respect to the orbital plane 
can account for
the whole phoenomenology of both SFXTs and persistently accreting HMXBs in general.
Both the orbital eccentricity and no-coplanarity can be explained 
by a substantial supernova kick 
at birth. This could indicate that SFXTs are likely young systems, probably younger than
persistent HMXBs. 

\acknowledgements Based on observations obtained with XMM-Newton, an ESA science
mission with instruments and contributions directly funded by ESA
member states and the USA (NASA).
Based on observations with INTEGRAL, an ESA project with instruments and the science data 
centre funded by ESA member states (especially the PI countries: 
Denmark, France, Germany, Italy, Switzerland, Spain), Czech Republic and Poland, 
and with the participation of Russia and the USA.
We thank the \xmm, \inte, and $Swift$ teams for making these observations possible,
in particular the duty scientists and science planners. PR thanks INAF-IASFMi for their kind hospitality. DG
acknowledges the French Space Agency (CNES) for financial support.
This work was supported by  contract ASI/INAF I/023/05/0.


\end{document}